\documentclass[twocolumn,showpacs,preprintnumbers,prb,amsmath,amssymb,floatfix,superscriptaddress]{revtex4} 
\usepackage{graphicx}
\usepackage{dcolumn}
\usepackage{bm}

\begin{document} 

\setlength{\topmargin}{0in}

\title{3-D mapping of diffuse scattering in 
Pb(Zn$_{1/3}$Nb$_{2/3}$)O$_3$-$x$PbTiO$_3$} 
\author{Guangyong Xu}
\affiliation{Brookhaven National Laboratory, Upton, New York 11973}
\author{Z. Zhong}
\affiliation{Brookhaven National Laboratory, Upton, New York 11973}
\author{H. Hiraka} 
\affiliation{Brookhaven National Laboratory, Upton, New York 11973}
\affiliation{Institute for Material Research, Tohoku University, 
Sendai 980-8577, Japan}
\author{G. Shirane}
\affiliation{Brookhaven National Laboratory, Upton, New York 11973}
\date{\today} 
 
\begin{abstract} 

High energy (67~keV) x-ray diffuse scattering measurements were performed 
on single crystals of Pb(Zn$_{1/3}$Nb$_{2/3}$)O$_3$-$x$PbTiO$_3$ (PZN-$x$PT).
A novel technique was developed to map out the diffuse scattering distribution 
in all three dimensions around a large number of Bragg peaks simultaneously, 
taking advantage of the almost flat Ewald sphere of the high energy x-ray beam. 
For $x=0, 4.5\%$, and $8\%$, the results are very similar, indicating 
same type of correlations of polarizations in these compounds. 
Our results show that the diffuse scattering intensity consists of  
six $\langle110\rangle$ rod-type intensities around reciprocal 
lattice points. A simple  
model is suggested where \{110\} type planar correlations of the in-plane  
$\langle1\bar{1}0\rangle$ type polarizations in real space contribute to 
the $\langle110\rangle$ 
rod-type diffuse intensities in the reciprocal space. The planar 
correlations of polarizations are likely a result from condensations of 
soft $\langle1\bar{1}0\rangle$ polarized optic phonon modes.

\end{abstract} 
 

\maketitle 

\section{Introduction}
Relaxors are a special class of ferroelectrics whose  dielectric properties
show diffusive and strongly frequency dependent phase transitions. 
Pb(Zn$_{1/3}$Nb$_{2/3}$)O$_3$ (PZN) is one of the prototype relaxors,
which has attracted much attention due to its high piezoelectric properties
when doped with PbTiO$_3$ (PT)~\cite{PZT1,PZN_phase1,PZN_phase2}. One 
unique property about relaxors is the appearing of local polarized 
nano-sized regions at the Burns temperature $T_d$, which is 
a few hundred degrees above the 
Curie temperature $T_C$.  This was first suggested by Burns and 
Dacol~\cite{Burns} in interpreting their measurements on the 
optical index of several relaxor systems, including PZN and one of 
its close analog, Pb(Mg$_{1/3}$Nb$_{2/3}$)O$_3$ (PMN). 

\begin{figure}[ht]
\includegraphics[width=\linewidth]{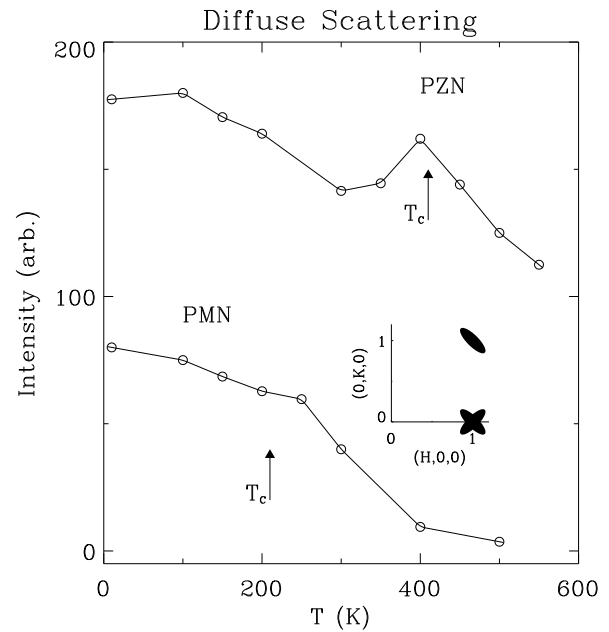} 
\caption{Neutron diffuse scattering intensity measured around (100) and 
(300) peaks for PMN and PZN, by Xu {\it et al.}~\cite{Xu_diffuse} and Stock 
{\it et al.}~\cite{Stock1}, respectively. The small anomaly in the PZN diffuse 
data around
$T=T_C$ is likely due to critical scattering at the phase transition. The 
inset shows schematics of diffuse scattering intensity distributions in the 
(HK0) plane, around the (100) and (110) Bragg peaks. }
\label{newfig}
\end{figure}

The PNR was then extensively studied by diffuse scattering 
measurements. It was found that in PMN, the diffuse scattering 
starts to appear at around $T_d \approx 600$~K, and 
increases monotonically
with cooling~\cite{PMN_neutron,PMN_neutron2,PMN_diffuse,Xu_diffuse}. 
Similar results were also observed in PZN~\cite{Stock1}
(see Fig.~\ref{newfig}), where $T_d$ is much higher than that of PMN. 
In order to understand the nature of polarizations and correlations/shapes of 
the PNR in relaxors, one of the most direct methods is to study the spatial 
distribution of the diffuse scattering intensity.  Recently, several
neutron~\cite{Xu_diffuse,Hiro_diffuse,PZN_diffuse,PZN_diffuse2,PZN_diffuse3} 
and x-ray~\cite{PMN_xraydiffuse,PMN_xraydiffuse2}
diffuse scattering measurements have been carried out on various relaxor 
compounds to investigate the shapes  of diffuse scattering intensity 
distributions in the reciprocal space around different Bragg 
peaks. 

However, the majority of
these measurements were performed in a 2-D scattering plane, i.e., probing the 
two-dimensional diffuse scattering distribution around Bragg peaks. 
For example, measurements for PMN around (100) and (110) Bragg peaks in the 
(HK0) scattering plane
(see the schematic representation in Fig.~\ref{newfig}) suggest that the 
diffuse scattering intensity extend along the $\langle110\rangle$ directions. 
But there is still ambiguity about  whether or not these are projections 
of diffuse scattering intensities along other out-of-plane  
directions (e.g., projection of [111] direction) onto the scattering plane.
You {\it et al.}~\cite{PMN_xraydiffuse} were the first to measure the 
out-of-plane x-ray diffuse scattering intensities of PMN around the (300) Bragg 
peak. Yet no systematic 3-D mapping of diffuse intensity distributions around 
different Bragg peaks in these relaxor systems has  been reported up to date. 
The lack of understanding of the diffuse scattering distribution in 
three-dimensions
made it very hard to establish an effective and convincing model that can 
explain and predict accurately the diffuse pattern.

In this paper, we present high energy x-ray diffraction results,
on the diffuse intensity distribution in three dimensions measured 
simultaneously around many Bragg peaks of PZN-$x$PT single crystals.
Our results show that the diffuse scattering intensities extend along  
all six $\langle110\rangle$ directions around most Bragg points. 
However, for certain Bragg peaks, one or more of the six $\langle110\rangle$ 
type diffuse intensities are absent, because of the associated polarization 
(atomic shift) being perpendicular to ${\bf Q}$. We suggest a simple 
phenomenological model, where $\langle1\bar{1}0\rangle$ polarizations are 
planar correlated in the \{110\} planes. 
For example, [1$\bar{1}$0] polarization correlated in the (110) 
plane. The diffuse scattering patterns calculated based on this simple model
are in good agreement with most previous and current results of 
measurements on diffuse scatterings from PNR in relaxors. 
Details of this novel technique, our results, and the model calculations, are 
discussed in the next three sections.

\section{Experimental Techniques}

Single crystals of PZN, PZN-4.5PT (4.5PT) and PZN-8PT (8PT) have been studied. 
The PZN single crystal is $3\times3\times1$~mm$^3$ is size, and  
was grown at the Simon Fraser University in Canada (the same
crystal used in Ref~\onlinecite{PZN_Xu}). The  4.5PT single crystal 
is $5\times5\times3$~mm$^3$ in size and provided by TRS ceramics.  
The 8PT crystal is a part ($\sim 4\times4\times1$~mm$^3$) 
of the  original crystal previously studied by Ohwada 
{\it et al.}~\cite{PZN_efield}, grown at the Pennsylvania State University. 

The x-ray diffraction measurements were performed at X17B1 beamline 
of the National Synchrotron Light Source (NSLS). 
A monochromatic x-ray beam of 67 keV, with an energy-resolution of
10$^{-4}$ ($\Delta$E/E), was
produced by a sagittal-focusing double-crystal monochromator 
using silicon [311] reflection with both crystals in asymmetric
Laue mode~\cite{zhong01_1}. In principle, with a four-circle 
x-ray diffractometer, one can reach any points in the reciprocal space, 
if allowed by geometry. However, in reality, most measurements 
on diffuse scatterings are performed in the ``zone''. In other words,
the measurements are performed mostly in a plane - the diffraction plane, 
around certain Bragg peaks. This is also the case for neutron diffuse 
scattering measurements. For example, measurements in the (HK0) zone indicate
that the measurements are performed in the plane perpendicular to the $c$-axis
(of the pseudocubic system), so that $L$ is fixed (to 0). This only probes two 
of the three dimensions of the 
diffuse scattering distribution around Bragg peaks. Certainly one 
can measure diffuse scatterings in other planes and essentially reconstruct a
3-D picture. But the beam time and efforts spent would usually be enormous.

\begin{figure}[ht]
\includegraphics[width=\linewidth]{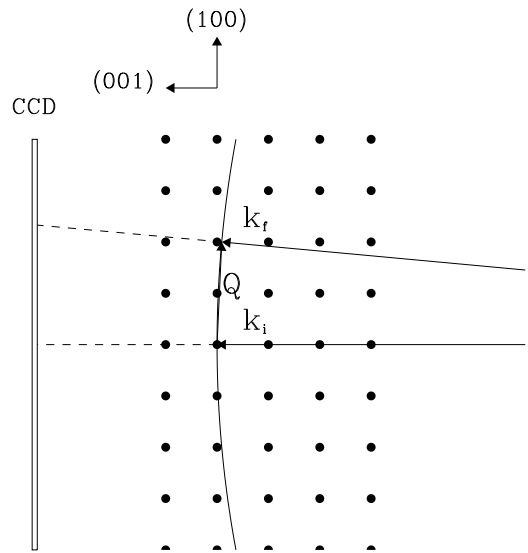} 
\caption{Schematics of the high energy x-ray diffuse measurement set-up.
The dots denote the reciprocal lattice points, and the arc describes where 
the Ewald sphere cuts in reciprocal space. The diffracted x-ray beam will then 
go into the CCD camera as shown after the sample.}
\label{fig0}
\end{figure}

Position sensitive detectors such as CCD detectors or Image-plates have 
been used to perform monochromatic Laue style measurements. In this type of 
measurements, scattering intensities in a plane, i.e., the Ewald sphere,  
can be taken simultaneously, very similar to the case of electron diffraction 
measurements in transmission electron microscopies. 
This usually requires Bragg points to be close to the Ewald sphere. 
In some cases, single crystal samples were oscillated while the image was taken.
This is a very efficient way to obtain 2-D diffraction intensity distributions, 
but without any information, or with very coarse resolution, along 
the incident beam direction perpendicular to the image plane. One example of 
this 
type of measurements on relaxor systems is shown in Ref.~\onlinecite{Pressure}.

In our measurements, we were able to take advantage
of the high energy (67~keV) of the x-ray beam and a position sensitive
CCD detector to perform 3-D x-ray diffuse scattering measurements in a 
very efficient manner. The focusing double-crystal monochromator set-up 
is also essential since it produces more than 
two orders of magnitude more photon flux than that of a conventional set-up.
A schematic of the experiment set-up is shown in Fig.~\ref{fig0}. The 
attenuation length of 67~keV x-rays in our samples are a few hundred microns, 
and the measurements are performed in a transmission mode. The 
reciprocal lattice and the radius of the Ewald sphere (the curve in 
Fig.~\ref{fig0}) are scaled from the real values in our measurements 
($a^*=2\pi/a=2\pi/4.06$~\AA~$=1.5476$~\AA$^{-1}$ and $k_i=33.87$~\AA$^{-1}$).
The incident x-ray beam was assumed to be along the reciprocal lattice [001] 
direction, and the [100] direction is pointing up.
Because of the high x-ray energy, and thus a large $k_i$, the Ewald sphere is 
almost flat, parallel to the (001) plane, perpendicular to $c$-axis 
at small ${\bf Q}$. 
This makes it possible to reach close to many reciprocal lattice points in 
the (HK0) zone 
simultaneously with one sample geometry.  In addition, it is important to 
note that at larger ${\bf Q}$ values, the Ewald sphere is not exactly cutting 
through the reciprocal lattice points.
Instead, the sphere cuts through the reciprocal lattice at  none-zero $L$ 
values, as well as a small tilt angle to the (001) plane. 
For example, around the (200) reciprocal point, the Ewald sphere is about 
$\delta L=0.09$~r.l.u. away from the exact (200) position. So we are in fact 
measuring 
diffraction intensities at (H,K,-0.09) around the (200) position on the CCD.
By tilting the sample, therefore tilting the reciprocal lattice, the 
Ewald sphere can cut the reciprocal lattice at different $\delta L$ values 
as we need. With a few measurements taken at different sample tilts,
the geometry of the diffuse scattering intensity distribution in three 
dimensions $I_{diff}(H,K,L)$ 
can be mapped out easily around selected Bragg peaks. Of course,
in our measurements, the Ewald sphere is curved not only vertically,
but also horizontally so that the situation would be very similar around 
the (020) peak as the (200) peak. 

Our measurements show that results from PZN, 4.5PT and 8PT crystals
are qualitatively the same. The x-ray beam size used in the PZN and 4.5PT 
measurements were $0.5\times0.5$~mm$^2$, and $0.2\times0.2$~mm$^2$ for the 
8PT measurements. The $q$-resolution of our measurements is mainly limited by 
the x-ray beam size.  The 8PT measurements with finer resolution show the 
features in a better detail, so in the next section we will focus on the 
results obtained using the 8PT sample.

\section{3-D x-ray diffuse scattering measurements}

\subsection{(HK0) zone}

\begin{figure*}[ht]
\includegraphics[width=\linewidth]{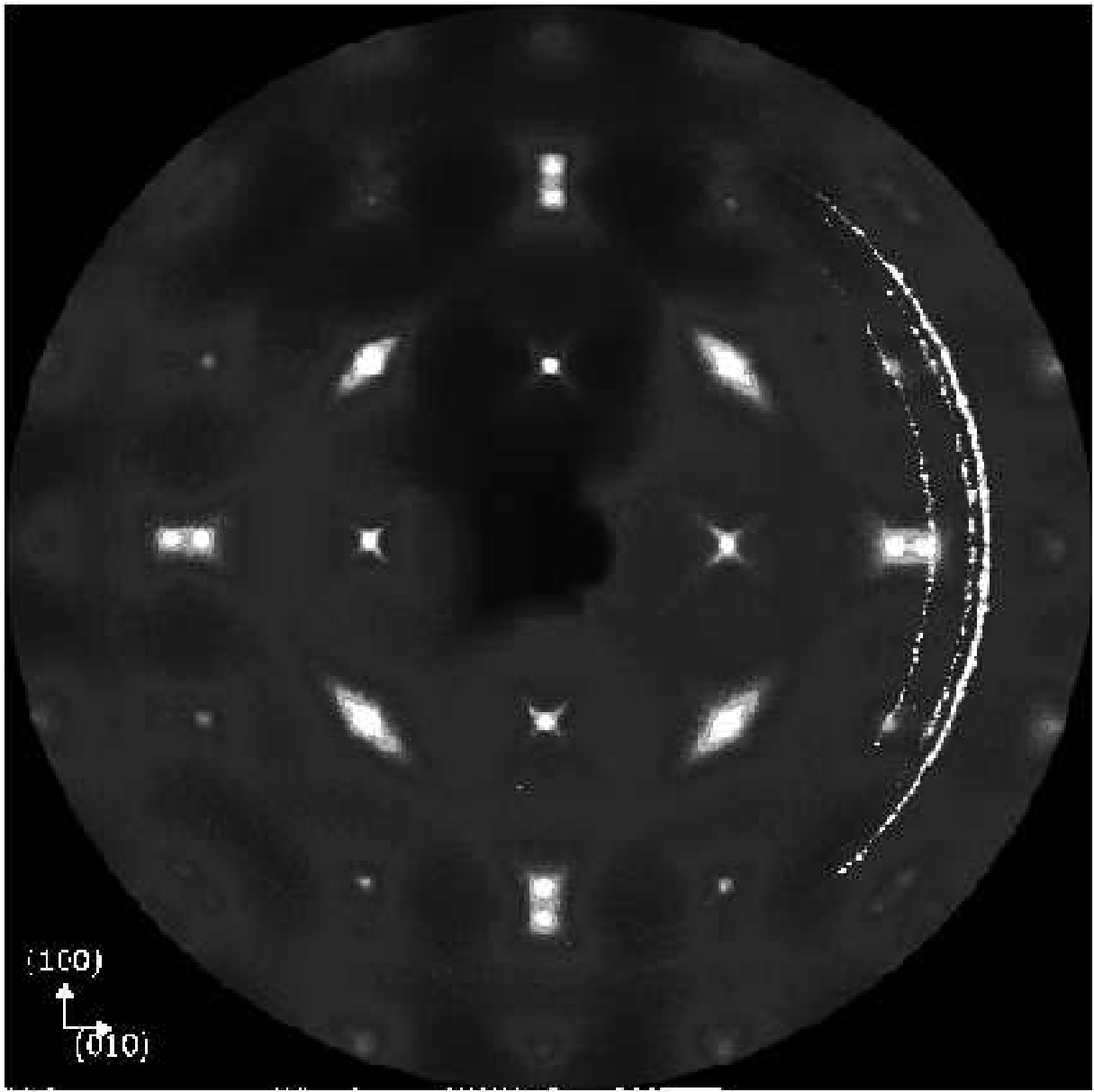} 
\caption{CCD image taken from PZN-8PT, at T=200~K. The incident x-ray beam
is along the [001] direction. The measurements are diffractions coming from 
very close to the reciprocal lattice (HK0) plane.}
\label{fig1_5}
\end{figure*}

Fig.~\ref{fig1_5} shows a CCD image taken at $T=200$~K, 
which is well below the ferroelectric transition temperature $T_C\sim 450$~K 
for PZN-8PT~\cite{Noheda,PZN_phase}. 
This was measured in the (HK0) zone, as shown by the schematics in 
Fig.~\ref{fig0}. The x-ray beam is incident in the [001] direction, 
perpendicular to the CCD plane. [100] and [010] directions are marked in the
figure. The center of the image, which is the origin of the reciprocal lattice, 
is black because a central beam stop was used 
to block the direct beam. The tilt (vertical) and rotation
(horizontal) of the sample has been aligned carefully so that the intensity
around (100), ($\bar{1}$00) are the same, as well as (010) and (0$\bar{1}$0), 
to ensure that the CCD plane is as parallel to the (001) plane as possible.
The counting time was 200 seconds, and we were able to measure very fine 
features simultaneously around many Bragg peaks for ${\bf Q}$ as large as 
(2,2,0). The rings in the image are powder rings from the Be window in the 
high temperature displex, and the rest of the background is mainly due to air 
scattering.

At the four \{100\} positions ((100), (010), ($\bar{1}$00), and (0$\bar{1}$0)), 
the Ewald sphere is still close to the Bragg
peak ($\delta L \sim 0.02$ r.l.u.). 
Tails of the Bragg peaks are picked up at these positions. In addition,
diffuse intensities extending out from the \{100\} Bragg peaks 
along 
the [110] and [1$\bar{1}$0] directions are clearly observed. Around the four 
\{110\} peaks ((110), (1$\bar{1}$0), ($\bar{1}$10), and ($\bar{1}\bar{1}$0)), 
we found that the diffuse scattering intensity is extending more in the 
transverse direction. For example, at the (110) peak, the diffuse scattering 
is strong along the [1$\bar{1}$0] direction. These $\langle110\rangle$ type
diffuse intensity distribution is in good agreement with previous 
x-ray~\cite{PMN_xraydiffuse,PMN_xraydiffuse2} and 
neutron~\cite{Xu_diffuse,Hiro_diffuse} diffuse measurements. 

When moving to
larger ${\bf Q}$ positions, the Ewald sphere starts to move further away from 
the (HK0) plane in the reciprocal space, and the results reveal more
interesting behavior. Around the (200) position, $\delta L \sim 0.09$ r.l.u., 
two spots are observed, split in the 
longitudinal direction along (100).  Taking into 
account the curvature of the Ewald sphere, the positions of the two spots
can be calculated to be (2.09,0,-0.09) and (1.91,0,-0.09).  One can see that
 (2.09,0,-0.09) $= 0.09\times$ (1,0,-1) $+$ (2,0,0), and 
(1.91,0,-0.09) $= -0.09\times$ (1,0,1) $+$ (2,0,0). 
It is possible that those could be diffuse intensities extending out 
from the (200) peak in the [101] and [10$\bar{1}$] directions. In addition, the
[110] and [1$\bar{1}$0] type diffuse intensities can still be vaguely seen
around the (200) peak. Although these diffuse intensities are much weaker 
than the Bragg peak intensity, they are also much broader in reciprocal space.  
This is why when the Ewald sphere is further away from the 
Bragg peak at (200) that that at (100), the sharp Bragg peak itself is not 
observed, but small traces of the [110] and [1$\bar{1}$0] type diffuse 
intensities still remain. 

\begin{figure*}[ht]
\includegraphics[width=\linewidth]{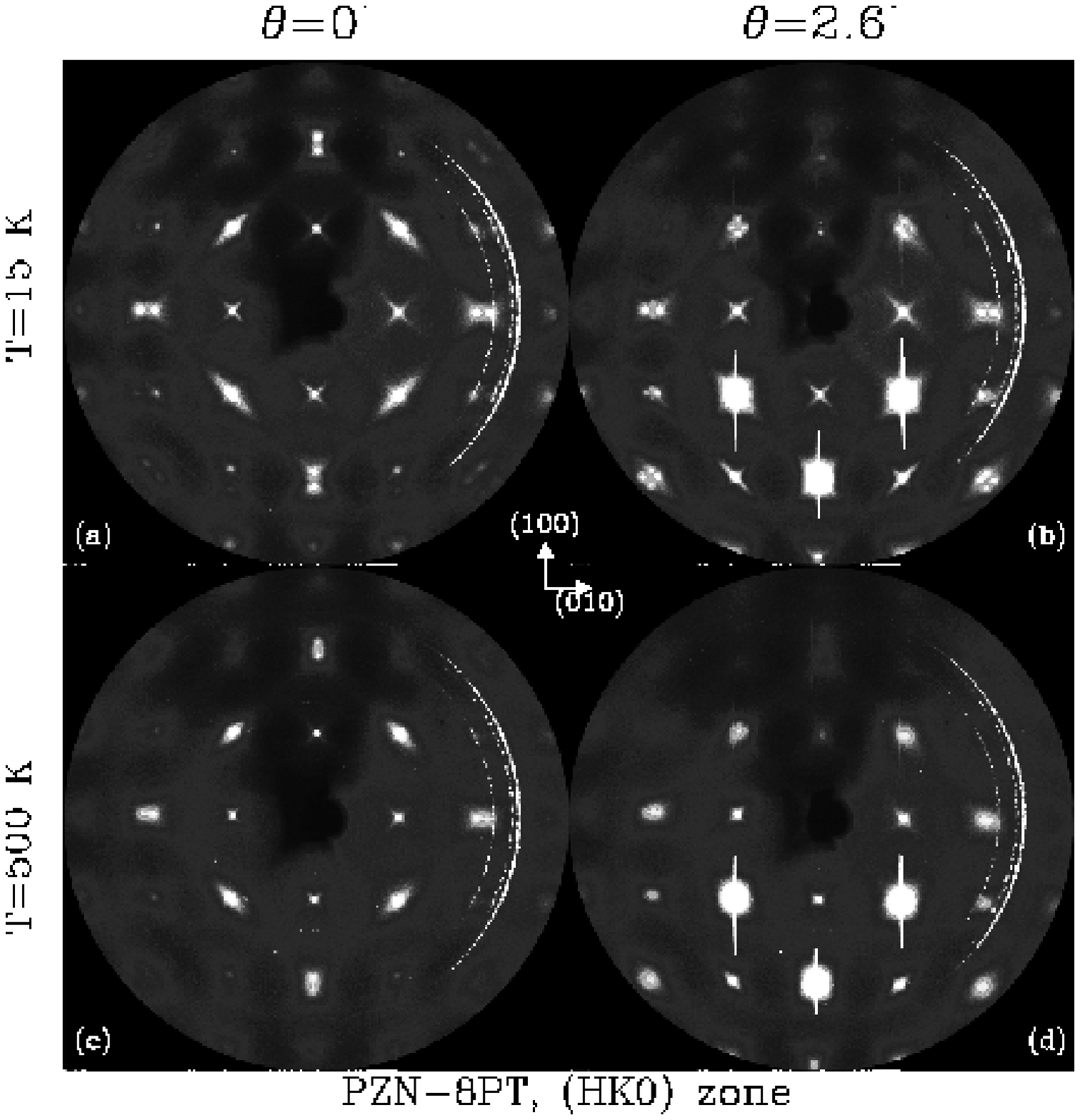} 
\caption{CCD image taken from PZN-8PT, at T=15~K and 500~K. The sample tilts are
$\theta=0^\circ$ and $2.6^\circ$, respectively. The incident x-ray beam
is along the [001] direction. The measurements are diffractions coming from 
very close to the reciprocal lattice (HK0) plane.}
\label{fig1}
\end{figure*}

Around the (210) position, in addition to the two spots split along (100) 
direction, we can see vaguely two weak spots split along the (010) direction 
too. This can be better seen at the (220) position, where the Ewald sphere 
is further away from the (HK0) plane, with $\delta L \sim 0.18$ r.l.u. 
Four spots are observed, positioned at around (2.18,2,-0.18), (1.82,2,-0.18), 
(2,2.18,-0.18), and (2,1.82,-0.18). Similar to the two spots 
around (200) position, these four spots are possibly intensity rods 
extending out 
from the (220) peak, along the [101], [10$\bar{1}$], [011], and [01$\bar{1}$]
directions, and cutting through the Ewald sphere with a none-zero L value.

\begin{figure*}[ht]
\includegraphics[width=\linewidth]{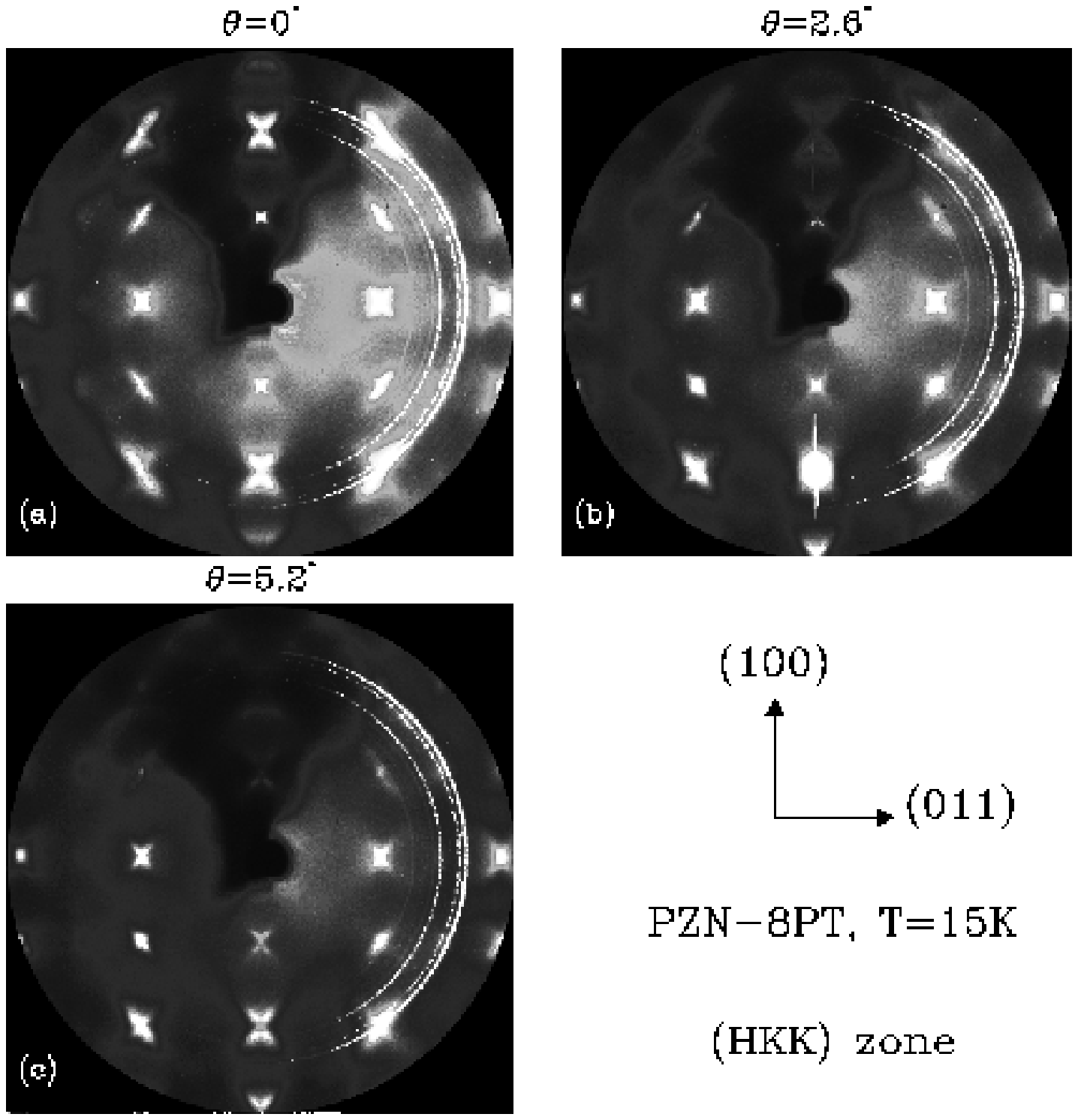} 
\caption{CCD image taken from PZN-8PT, at T=15~K. The sample tilts are
$\theta=0^\circ$, $2.6^\circ$, and $5.2^\circ$, respectively. 
The incident x-ray beam
is along the [01$\bar{1}$] direction. The measurements are diffractions coming 
from very close to the reciprocal lattice (HKK) plane.}
\label{fig2}
\end{figure*}

In order to verify this, more measurements were performed by tilting the sample,
so that diffuse scattering intensities at different $\delta L$ values away from 
the Bragg peaks can be studied.
In Fig.~\ref{fig1}, images taken at $T=15$~K and 500~K, as well as different 
sample tilt angles are shown. At $T=15$~K, sample tilt $\theta=0^\circ$, 
the result (Fig.~\ref{fig1} (a)) looks very similar to that at $T=200$~K, 
shown in Fig.~\ref{fig1_5}.
When the sample was tilted by 2.6$^\circ$ (Fig.~\ref{fig1} (b)), 
the ($\bar{2}$00) Bragg peak was 
actually right on the Ewald sphere, and the CCD was saturated at that position. 
The vertical lines in the image are artifacts of the CCD saturation. The 
(200) peak, however, was moved further away from the Ewald sphere. The sphere 
is therefore cutting at a larger $\delta L \sim 0.18$ around the (200) 
position. The 
two spots around (200) on the image clearly split further apart than that 
measured without sample tilt (Fig.~\ref{fig1} (a)). They are roughly positioned 
at (2.18,0,-0.18) and (1.82,0,-0.18). This confirms that these two 
spots are indeed results from Ewald sphere cutting the intensity ``rods'' 
coming out from the (200) Bragg peak along the [101] and [10$\bar{1}$] 
directions. 
The sample tilt also moved the (100), (110) and (1$\bar{1}$0) peaks (on the 
top part of the image) further away from the Ewald sphere. 
The fine features around these positions are 
also consistent with those around (200) and (220). Two longitudinally split 
spots at (100), and four spots at (110) and (1$\bar{1}$0). In 
Fig.~\ref{fig1} (a), these Bragg points are simply too close to the Ewald 
sphere for any fine features to be observed. The ($\bar{2}$20) 
and ($\bar{2}\bar{2}$0) peaks (on the bottom part of the image) are, however, 
moved closer to the Ewald sphere in 
Fig.~\ref{fig1} (b) then in (a). Here we see the four spots moving closer in,
also in good agreement with what one would expect from the Ewald sphere cutting
intensity ``rods'' along [101], [10$\bar{1}$], [011], and [01$\bar{1}$]
directions.

One important fact to note is that around the (200) position, no intensity 
rods along the [011] and [01$\bar{1}$] was observed. Similarly, around
the (020) position, no intensity rods along the [101] and [10$\bar{1}$] 
directions are present. In x-ray and neutron scattering measurements, the 
diffuse scattering intensity resulting from correlated 
polarizations (atomic shifts) 
is proportional to $|{\bf Q\cdot\epsilon}|^2$, where ${\bf \epsilon}$ is the 
polarization vector. It is therefore evidential that the polarizations
contributing to the [011] and [01$\bar{1}$] type diffuse scatterings 
are perpendicular to ${\bf Q}=$(1,0,0).

\subsection{(HKK) zone}

Diffuse scattering measurements were also performed in a different 
zone, the (HKK) zone. In Fig.~\ref{fig2}, results are shown at $T=15$~K,
with the sample rotated by 45$^\circ$ around the vertical 
axis ([100] direction). Now the x-ray beam is incident in the [01$\bar{1}$]
direction, and the Ewald sphere is almost parallel to the reciprocal 
(HKK) plane, i.e., the plane defined by the [100] and [011] vectors. 

With sample tilt $\theta=0^\circ$, we see four spots around the (200) position 
as expected, where the Ewald sphere is about $\delta q=$(0,-0.064,0.064) away
from the (HKK) plane. These four spots are roughly positioned at 
(2.13,0,0.13), (2.13,-0.13,0), (1.87,0,0.13), and (1.87,-0.13,0). 
It is entirely consistent with the results from the (HK0) zone measurements,
i.e., intensity rods extending out in the [101], [10$\bar{1}$], [110], 
and [1$\bar{1}$0] directions from the (200) peak, being cut by the Ewald 
sphere. With tilt $\theta=2.6^\circ$ (see Fig.~\ref{fig2} (b)), these four 
spots move further out, as the Ewald sphere is moving further away from the 
(HKK) plane around (200). No central peak is seen at the center of those four
spots (${\bf Q}=$(2.,-0.64,0.64)), confirming the absence of diffuse intensity 
rods in the [01$\bar{1}$] direction around the (200) position. 
In addition, one starts to see similar features 
around the (100) position as well. With further tilting, $\theta=5.2^\circ$
(see Fig.~\ref{fig2} (c)), the intensities around (200) has become very weak, 
and the further parted four spots can hardly be observed. On the bottom of the 
image, four spots also appear around the ($\bar{2}$00) position. There the 
Ewald sphere actually went passed the (HKK) plane, with 
$\delta q=$(0,0.064,-0.064), so similar patterns are seen there compared to 
Fig.~\ref{fig2} (a).

Diffuse patterns around other Bragg positions are all consistent with 
this $\langle110\rangle$ type diffuse rods. For example, if we look at 
around the (220) position, a strong peak appears there.  
The Bragg structure factor of PZN (220) and (200) are the same order of 
magnitude, yet the Ewald sphere is further away from the Bragg peak at (220) 
than (200) (because of the larger ${\bf Q}$). Since the (200) Bragg is not 
observed,
neither should the (220) Bragg. In fact, that ``peak'' is not 
the (220) Bragg peak, but instead the diffuse scattering intensity coming out 
from the (220) peak along the [0$\bar{1}$1] direction (perpendicular to the 
image plane). 

\subsection{3-D diffuse intensity distribution}

\begin{figure}[ht]
\includegraphics[width=\linewidth]{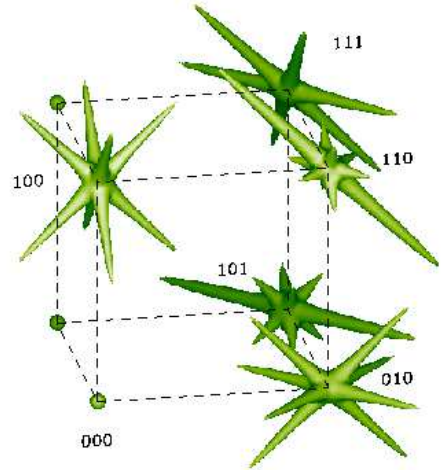} 
\caption{Sketch of the diffuse scattering distribution in 
the 3-D reciprocal space around (100), (110), (111), (010), and (011) 
reciprocal lattice points. 
}
\label{fig3_0}
\end{figure}

Our results can establish  unambiguously  the geometry of the diffuse 
scattering 
intensities distributed in the reciprocal space.  Based on all the information
obtained, (i) diffuse intensities are $\langle110\rangle$ 
type rods around Bragg peaks; (ii) the intensities of different 
$\langle110\rangle$ 
diffuse rods can vary between different Bragg reflections; 
(iii) at some Bragg reflections, certain $\langle110\rangle$ intensity 
rods are strong, and certain $\langle110\rangle$ rods may even be absent, 
the 3-D distribution of the diffuse scattering
intensity can be sketched in Fig.~\ref{fig3_0}. The 3-D sketch is shown in the 
way that the (001) plane ((HK0) plane) is parallel to the paper. Here we 
plot the equi-intensity surface of the diffuse scattering around various
Bragg peaks. The contrast and sharpness
of the diffuse intensity rods have been exaggerated in order to illustrate
the geometry in a simple and effective way.   

In a quantitative way, it 
describes our results very well. For example,  around the (100) peak, 
the [110], [1$\bar{1}$0], [101], and [10$\bar{1}$] diffuse rods are of 
equal intensities, but 
the [011] and [01$\bar{1}$] rods are absent; around the (110) peak, the 
[1$\bar{1}$0] diffuse rod is much more intense than the [101], [10$\bar{1}$], 
[011], and [01$\bar{1}$] rods, but the [110] rod is absent. 
A cross section of the 3-D diffuse scattering distribution in the (HK0) 
plane gives the ``butter-fly'' shaped pattern around the (100) peak,
and the transverse [1$\bar{1}$0] type diffuse around the (110) peak, 
in good agreements with previous neutron scattering 
measurements~\cite{Xu_diffuse,Hiro_diffuse}. 
More detailed model calculations will be presented in the next section.

\section{Discussions}

\subsection{Static or dynamic?}

\begin{figure}[ht]
\includegraphics[width=\linewidth]{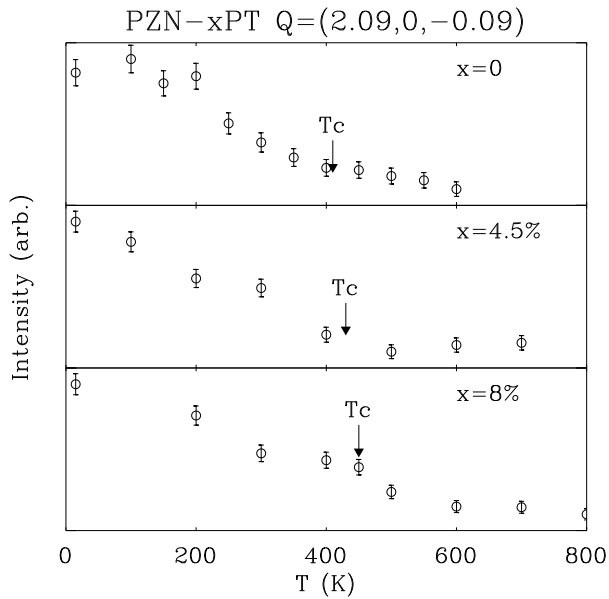} 
\caption{Temperature dependence of the diffuse scattering measured at 
(2.09,0,-0.09) for PZN, PZN-4.5PT and PZN-8PT.}
\label{fig4}
\end{figure}

One disadvantage of x-ray diffraction measurements is the 
coarse energy resolution, so that many of the low energy phonon modes
are also included together with the elastic component.
Some examples of phonon contributions dominating the x-ray diffuse intensity 
distribution can be found in similar ferroelectric perovskite compounds 
BaTiO$_3$~\cite{BaTiO3} and KTaO$_3$~\cite{KTaO3}.
A very important question to resolve is whether these intensity rods observed 
in PZN-$x$PT diffuse  are coming from 
phonon contributions (thermal diffuse) or static distortions (static diffuse).

Recently,  neutron diffuse scattering 
measurements~\cite{Xu_diffuse,Hiro_diffuse,PZN_diffuse3,PMN_diffuse3} 
have been performed 
on the relaxor systems with good energy 
resolutions so that phonon intensities can be easily separated out. These
results  
have directly confirmed that the $\langle110\rangle$ type  diffuse intensities 
in the (HK0) plane are elastic.

The temperature dependence of the diffuse scattering intensity is also 
consistent with the static nature.
In Fig.~\ref{fig4}, the temperature dependence of the 
diffuse scattering intensity at (2.09,0,-0.09) (the top one of the two spots
observed around the (200) position) are plotted for PZN, 4.5PT, and 8PT.
At T below $T_C$, the intensity decreases slowly and monotonically with 
increasing T for all three compounds. In addition, the diffuse intensity 
does not reach zero at the ferroelectric phase transition, but rather at 
some temperatures above $T_C$. All these are in agreement with the 
elastic diffuse scatterings expected from PNR in relaxor ferroelectric systems.

Phonon scattering intensities, on the other hand, are proportional
to $k_BT/(\hbar\omega)^2$, when taking into account of the 
Bose factor  $k_BT/\hbar\omega$ and the $1/\omega$ factor. 
In PZN~\cite{Stock1} and 
PMN~\cite{Waki1}, the zone center TO phonon energy actually increases with 
cooling for $T<T_C$. Measurements
on TA phonons at small $q$ in the same systems~\cite{Waki1,Stock1} 
also confirmed that the TA 
phonon modes behave normally below $T_C$, i.e., no significant 
broadening or softening. Therefore, both TO and TA phonon contribution to 
the x-ray diffuse scattering should increase
with increasing T, at least for the low temperature range $T<T_C$. This is
in contrast to the data shown in Fig.~\ref{fig4}. As a comparison, 
x-ray diffuse 
scattering measurements with similar set-ups on single crystals of Si have 
been performed by us, and by 
Holt {\it et al.}~\cite{Si_Phonon}. There the phonon contribution dominates, and
the thermal diffuse intensity clearly increases with heating.

\begin{table*}[ht]
\caption{X-ray scattering structure factors for PZN. 
The diffuse scattering structure 
factors are based on the atomic shift values derived 
from measurements by Vakhrushev {\it et al.}~\cite{PMN_neutron3} on PMN. The
soft optic phonon structure factors are calculated based on a ratio of 
$S=1.5$ between
the Last Mode and Slater mode contributions, as determined by Hirota 
{\it et al.}~\cite{PMN_diffuse}}
\begin{ruledtabular}
\begin{tabular}{lcccccccc}
&(100)&(110)&(111)&(200)&(210)&(220)&(222)&(300)\\
$Q^2|F_{Bragg}({\bf Q})|^2$ &1&19&11&63&5&126&190&9\\
$Q^2|F_{Diff}({\bf Q})|^2$&10&27&17&33&50&66&99&90\\
$Q^2|F_{Soft}({\bf Q})|^2$&88&43&30&10&439&20&30&789

\end{tabular}
\end{ruledtabular}
\label{tab1}
\end{table*}

Another clue is provided by  comparing the structure factors
calculated based on previous results. Previous measurements by You 
{\it et al.} obtained similar results
near the (300) Bragg peak. It was attributed to contributions from
ferroelectric soft transverse optic phonons propagating along the 
$\langle110\rangle$
directions, and thus concluded to be only strong around the odd integer
Bragg reflections ($H+K+L=$~odd). In our measurements, however, 
the diffuse scattering intensities around the 
(200), (110), and (220) Bragg peaks are reasonably strong. 
The static diffuse scattering, acoustic and  and 
soft optic phonon structure factors at different Bragg reflections are 
shown in Table.~\ref{tab1}. Apparently the diffuse scattering 
intensities observed in our measurements are more consistent with the static 
diffuse structure factors, which do not vary much between different Bragg peaks 
in our measurement range.  It is also qualitatively in agreement with 
measurements by Takesue {\it et al.}~\cite{PMN_xraydiffuse2} on PMN in the 
(HK0) plane. 
The soft phonon structure factors, however, 
vary a lot and are very strong at (210) and (300). 
Measurements around those peaks did not show any particularly strong diffuse 
intensities (see Figs.~\ref{fig1_5} and \ref{fig1}). 

These facts strongly suggest that for temperatures $T<T_C$, 
the main contribution 
to our x-ray diffuse scattering is coming from static lattice distortions, 
namely, those from the polarized PNR, instead of thermally activated 
inelastic phonon scatterings.  
We can therefore study the low temperature diffuse scattering 
distributions to better understand the static polarizations (atomic shifts)
and correlations 
in the PNR. With increasing temperature, the phonon contributions will 
eventually increase and dominate. In fact, at 700~K, with the disappearance of 
the two spots around (200) position, we started to pick up intensities 
around (2,0,-0.09), which is very likely coming from thermally activated 
phonons propagating in the [001] direction.

In the next two subsections, we will start at providing a simple phenomenological 
model describing the real space correlation of polarizations,
that gives rise to the $\langle110\rangle$ rod-type diffuse intensities.  
Based on this model, the shape of diffuse scattering in these compounds can
be well calculated within good agreement to the experimental results.  
We will then  suggest one possible origin of these elastic diffuse 
scatterings from the aspect of lattice dynamics - condensation of 
soft optic phonons into static distortions.

\subsection{Planar correlations - ``pancake'' model}

\begin{figure}[ht]
\includegraphics[width=\linewidth]{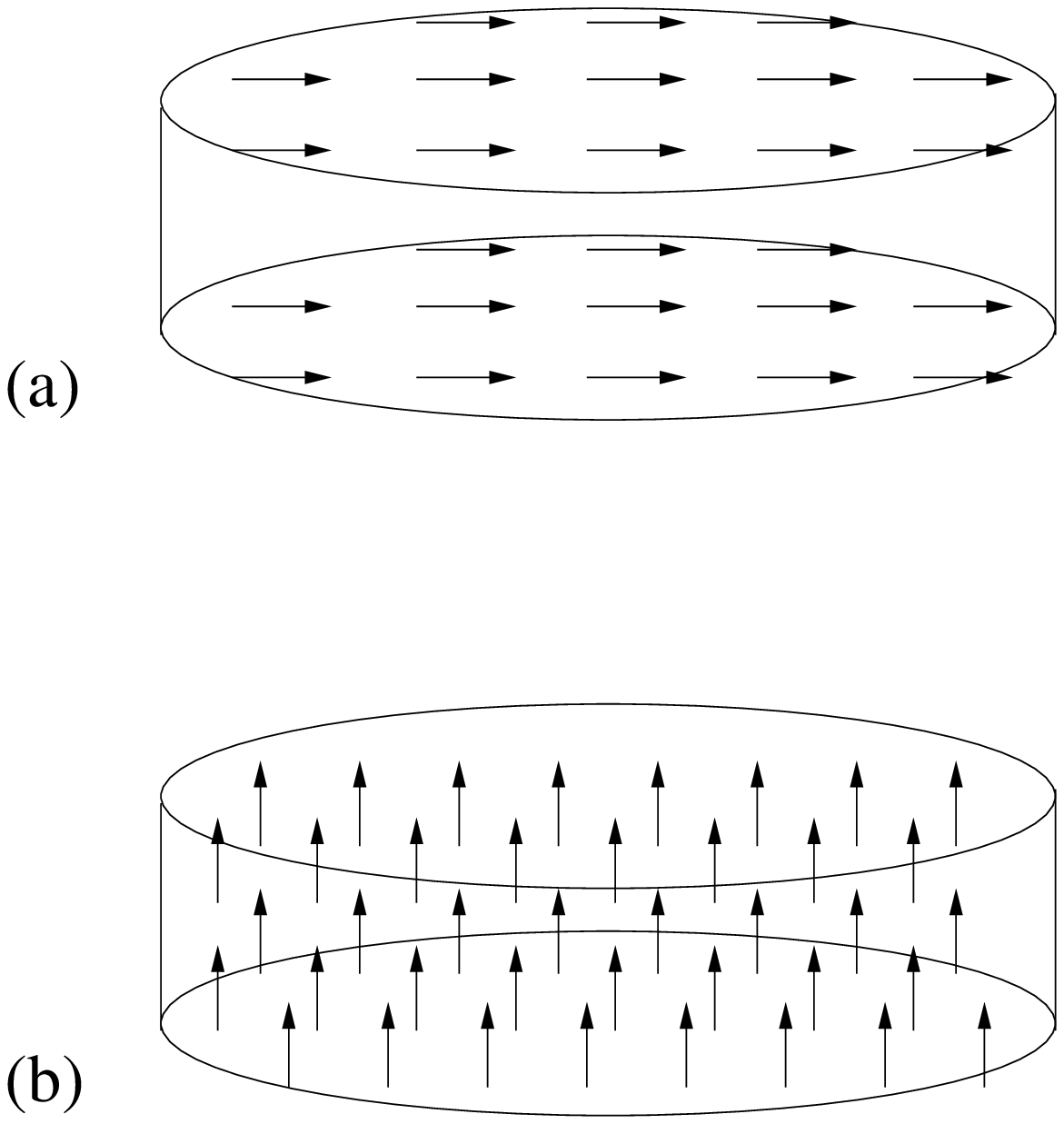}
\caption{Polarizations confined in the ``pancake'' shaped spaces. The 
correlation in the plane (diameter) is much larger than the correlation 
perpendicular to the plane (thickness). (a) and (b) are two different models 
where the polarizations are in the plane, or out of the plane, respectively.}
\label{fig_pancake}
\end{figure}

Having determined the static nature of these observed $\langle110\rangle$ 
type diffuse intensity rods, we now consider a simple model that can 
reproduce such diffuse intensities. Generally, a rod type structure in 
the reciprocal space corresponds to a planar structure in the real space.
Therefore, the $\langle110\rangle$ type diffuse scattering intensity can 
be a result of polarizations correlated in the 
\{110\} planes in the real space. There are totally six \{110\} planes, 
corresponds to the six $\langle110\rangle$ diffuse intensity rods. 
Furthermore, the polarizations should have the same periodicity as the 
lattice itself, because all these diffuse scattering intensities 
peak at the Bragg positions, instead of forming some super-lattice peaks.

\begin{figure}[ht]
\includegraphics[width=\linewidth]{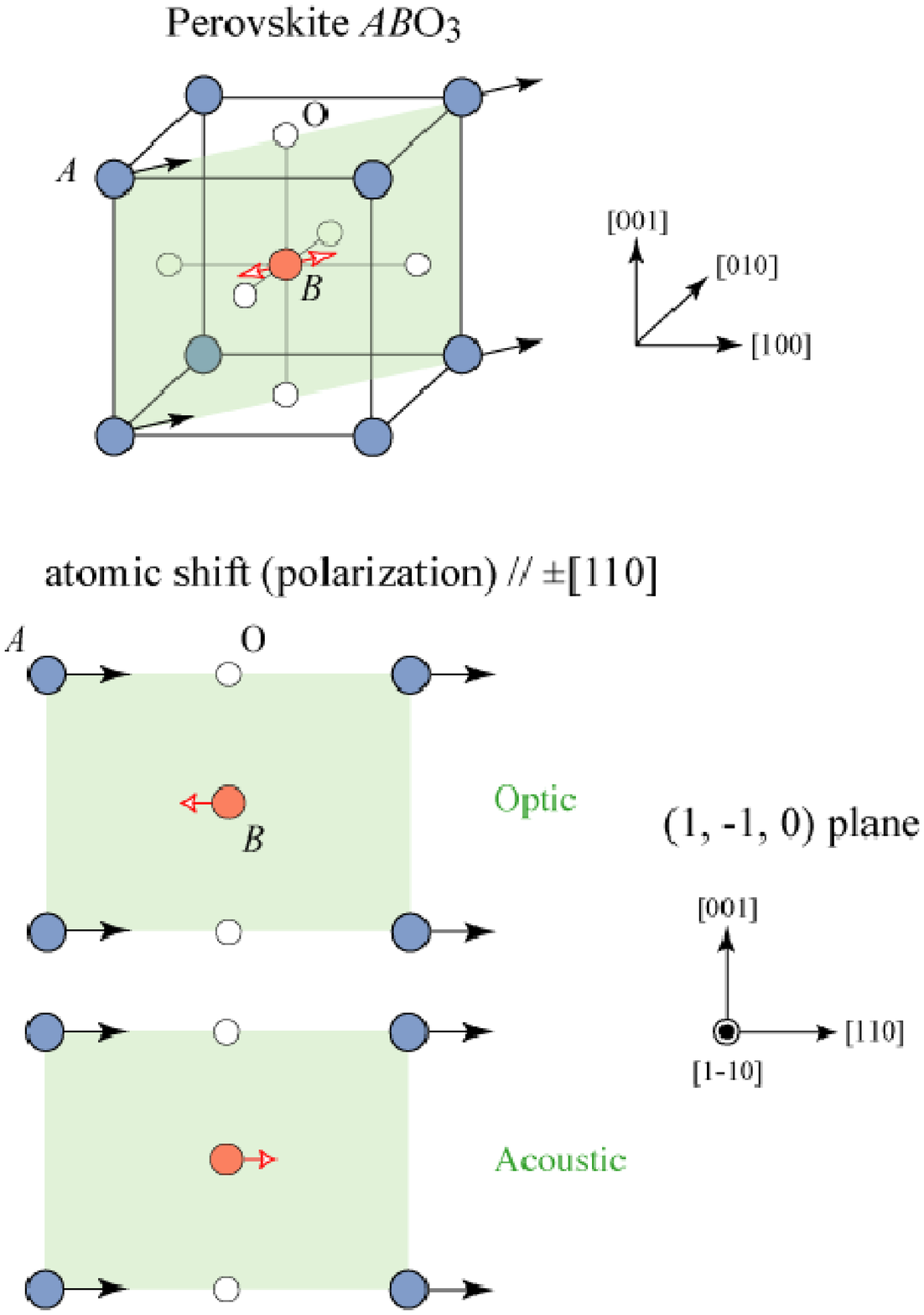}
\caption{Real space model indicating the correlation of the  polarizations
(polarizations) in the system. Here the [110]  polarizations 
are correlated in the (1$\bar{1}$0) plane. }
\label{fig5}
\end{figure}

The problem remaining is to determine the directions of these polarizations
with planar type correlations. The polarizations can be either 
in-plane,  or out-of-plane, as shown in Fig.~\ref{fig_pancake}. Here 
the polarizations are confined in the ``pancake'' shaped spaces, 
indicating the planar type correlations. One can easily find that 
out-of-plane type polarizations as shown in Fig.~\ref{fig_pancake} (b) do not
yield the type of diffuse intensity pattern around the (110) peak. There when 
measured in the (HK0) plane, the diffuse intensity is mostly transverse along
the [1$\bar{1}$0] direction, and the [110] type intensity rod is absent.
Since $I_{diff}\propto |{\bf Q\cdot\epsilon}|^2$, polarizations $\epsilon$ 
correlated in the (110) plane, that give rise to the [110] type diffuse,  
must be perpendicular to ${\bf Q}=$(1,1,0) for this intensity to be 
zero. Only in-plane type polarizations can satisfy this condition. 

Nevertheless, the polarizations still can not rotate freely in the \{110\} 
planes. The exact direction of these polarizations can be 
determined by studying the diffuse intensity around the (200) peak. Here
the diffuse scatterings along [011] and [01$\bar{1}$] directions are absent.
So, for the [011] type diffuse, $\epsilon$ must be perpendicular to 
both ${\bf Q}=$(2,0,0) (so that ${\bf Q\cdot\epsilon}=0$) and the out-of-plane 
vector [011] (so that $\epsilon$ is in-plane). This
immediately leads to $\epsilon=$[01$\bar{1}$]. Similar analysis can 
be used to derive the polarizations directions associated to all the 
six $\langle110\rangle$ type diffuse scatterings. 

In summary, we find that the real space structure corresponding to the 
$\langle110\rangle$ rod-type diffuse intensity is \{110\} planar
correlations of in-plane $\langle1\bar{1}0\rangle$ polarizations.
More specifically, there are polarizations along the [110] direction, correlated
in the (1$\bar{1}$0) plane; polarizations along the [1$\bar{1}$0] direction, 
correlated in the (110) plane; polarizations along the [101] direction, 
correlated
in the (10$\bar{1}$) plane; polarizations along the [10$\bar{1}$] direction, 
correlated in the (101) plane; polarizations along the [011] direction, 
correlated
in the (01$\bar{1}$) plane; and polarizations along the [01$\bar{1}$] 
direction, correlated in the (011) plane. Fig.~\ref{fig5} provides a schematic 
illustration of one of these six planar correlations in real space. 
The x-ray diffraction measurements are less sensitive to the lighter atoms 
such as O, so we only illustrated possible motions of the heavier 
A (Pb$^{2+}$) and B (Zn$^{2+}$/Nb$^{5+}$/Ti$^{4+}$) site atoms.
The motions of the A site and B  site can be parallel (acoustic) or 
antiparallel (optic), or a combination of the two.

\begin{figure}[ht]
\includegraphics[width=\linewidth]{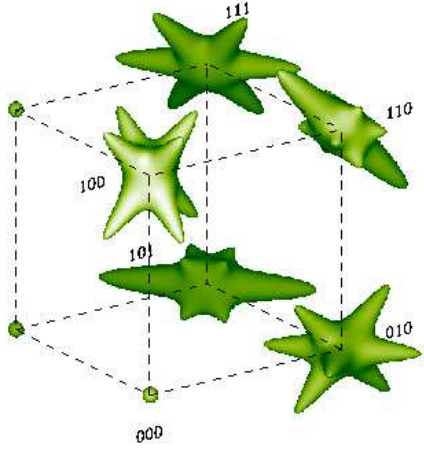} 
\caption{Model calculation of the diffuse scattering distribution in 
the 3-D reciprocal space around (100), (110), (111), (010), and (011) 
reciprocal lattice points.}
\label{fig3}
\end{figure}

The diffuse scattering intensity can be therefore calculated based on this 
simple model. At ${\bf Q}={\bf G}+{\bf q}$, 
\begin{equation}
I_{diff}({\bf Q})=A|F_{diff}({\bf G})|^2\sum_i|{\bf Q\cdot\epsilon_i}|^2
\frac{\Gamma_\parallel}{q_{\parallel i}^2+\Gamma_\parallel^2}
\cdot\frac{\Gamma_\perp}{q_{\perp i}^2+\Gamma_\perp^2},
\end{equation} 
$I_{diff}({\bf Q})$ is the sum of contributions from the six \{110\} 
planar type correlations, where $|F_{diff}({\bf G})|^2$ is the diffuse 
scattering structure
factor at the  Bragg reflection ${\bf G}$; $\epsilon_i$ is the polarization
vector correlated in the plane (one of the six $\langle110\rangle$); 
$\Gamma_\parallel=1/\xi_\parallel$ and $\Gamma_\perp=1/\xi_\perp$, are the 
inverse of the correlation length in- and out-of-plane;  $q_{\parallel i}$
and $q_{\perp i}$ are the in- and out-of-plane components of ${\bf q}$. Here 
we used a product of Lorentzian functions because 
an exponentially decaying correlation in real space, both in- and out-of-plane, 
has been assumed.
For a square-function cut-off type correlation caused by a finite ``pancake'' 
shape, with thickness described by $\xi_{\perp}$ and diameter described 
by $\xi_{\parallel}$, 
the reciprocal space intensity distribution will take the functional forms of 
Gaussian functions instead. However, the shape of the diffuse scattering 
intensity will not change qualitatively.

In the simulation, the in-plane correlation length was 
assumed as 20 lattices, 4 times larger than  the out-of-plane correlation length
(5 lattices - estimated from results of neutron diffuse measurements on PMN by 
Xu {\it et al.}~\cite{Xu_diffuse}).
The contributions from the six \{110\} planar correlations are weighed
by the $|{\bf Q\cdot\epsilon}|^2$ factor and summed up. The result is shown
in Fig.~\ref{fig3}. Here the equi-intensity 
surface of the diffuse scattering intensity
is plotted. Note that for best visual effects, the static diffuse structure 
factors in Table.~\ref{tab1} and the $|{\bf Q}|^2$ 
factor were taken out in constructing Fig.~\ref{fig3}, in order to 
illustrate the diffuse intensity distributions around different Bragg 
peaks with the same quality in the same figure. In the 
exaggerated version of this model, as shown in Fig.~\ref{fig3_0}, the 
in-plane correlation length was chosen to be 20 times larger than the 
out-of-plane one to produce those ``sharp'' diffuse rods.

\begin{figure}[ht]
\includegraphics[width=0.8\linewidth]{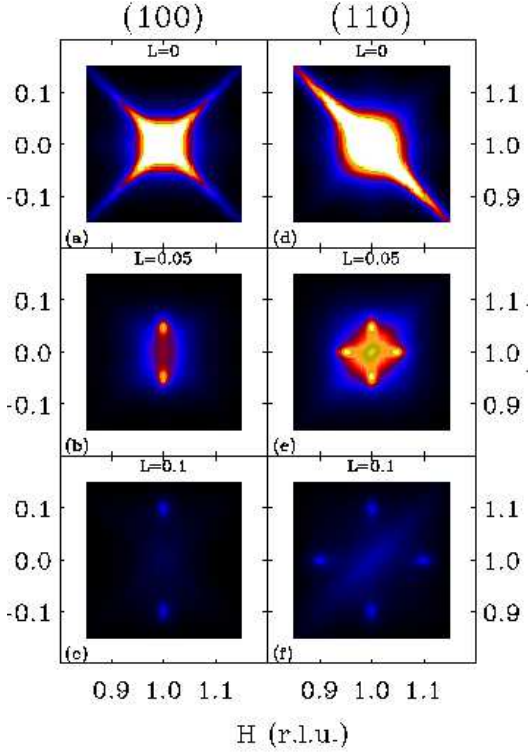} 
\caption{Model calculation of the diffuse scattering distribution in 
the (H,K,0), (H,K,0.05) and (H,K,0.1) reciprocal planes (i.e. (001) planes, at 
different small L values), around the (100) and (110) reciprocal lattice points.
}
\label{fighk0}
\end{figure}

In Fig.~\ref{fig3}, the simulated diffuse scattering intensities around the 
(100), (010), (110), (111), and (011) Bragg peaks are plotted. When measured
in the (HK0) plane, the cross section of the 3-D diffuse scattering intensities
gives the ``butter-fly'' diffuse pattern around (100) 
and (010) peaks, as well as the transverse [1$\bar{1}$0] type diffuse 
around the (110) peak. By moving away from the (HK0) plane, with a small 
$\delta L$, one starts to see the two spots split in the longitudinal direction
around the (100) Bragg peak due to the [101] and [10$\bar{1}$] type diffuse 
intensities; and four spots around the (110) Bragg peak due to the [101], 
[10$\bar{1}$], [011], and [01$\bar{1}$] type diffuse intensities. A simulation 
of diffuse intensities in the (HK$\delta L$) plane, 
at different $\delta L$ values is shown in Fig.~\ref{fighk0}. It is 
in good agreement with our measurements in the (HK0) zone. Similarly, one can
also compare the simulation in the (HKK) zone to our measurements.
We found that calculations based on this model can explain all of our 
results, as well as almost all 
of the previous results from x-ray~\cite{PMN_xraydiffuse, PMN_xraydiffuse2} and 
neutron~\cite{Xu_diffuse,Hiro_diffuse,PZN_diffuse3} diffuse scattering 
measurements on PMN and PZN-$x$PT. Although it does not yet provide a 
clear physical picture on the origin of these correlated polarizations, it 
is very useful in calculating and predicting diffuse scattering distributions
in these relaxor compounds.

One interesting problem is that all of the three compounds
in our measurements have rhombohedral type polarizations below $T_C$. 
Recent NMR measurements~\cite{PMN_nmr} on PMN also confirms the 
$\langle111\rangle$ type polarization in the system. How do the 
$\langle111\rangle$ type polarization and the correlated $\langle110\rangle$ 
type polarization coexist in the same system? In fact, our model is not in 
contradiction to the average $\langle111\rangle$ type polarization. 
Alternatively, the $\langle111\rangle$ type polarization can be decomposed into 
three $\langle110\rangle$ components. The average 
$\langle111\rangle$ type polarization can be a result of global averaging
of $\langle110\rangle$ type polarized ``pancake'' entities.  
Another possible scenario  could be the following: when three different
\{110\} type ``pancakes'' (e.g., (110), (011), and (101)) cross each other,
the region shared by those three ``pancakes'' would have a combination 
of three $\langle110\rangle$ polarizations - resulting in 
a $\langle111\rangle$ type polarization. This shared region could be a
$\langle111\rangle$ polarized PNR, and the size of this region
is then defined by the thickness of the three ``pancakes'', i.e., 
the out-of-plane correlation length, which is in the order of one to few 
nanometers (a few lattice units).

\subsection{Condensation of $\langle1\bar{1}0\rangle$ polarized soft optic 
phonon modes}

Ferroelectric polarizations are optic type distortions, associated with 
different atoms moving opposite to each other in the unit cell.
It is well accepted that the PNR are formed from the condensation of the 
ferroelectric soft optic phonon mode. Previous measurements show that the 
transverse optic (TO) phonons become overdamped for $T_C < T < T_d$ for $q$ 
smaller than a certain value, called the ``waterfall'' wave-vector 
$q_{wf}$~\cite{PZN_waterfall1,PZN_waterfall2}. The ``water-fall'' effect
was observed for $q$ along both [001] and [011] 
directions~\cite{PZN_waterfall1}, and 
is believed to be intimately related to some length scales defined by the 
PNR. We can therefore suggest one possible origin of the 
rod-type diffuse scattering intensities. 
If the $\langle1\bar{1}0\rangle$ polarized optic phonon modes are 
particularly soft when propagating along the perpendicular $\langle110\rangle$
directions, they can then condense into the $\langle110\rangle$ type 
elastic diffuse intensities upon cooling. This occurs around $T=T_d$,
when the soft optic mode becomes overdamped and start to condense into static 
distortions. The shape of these static entities are then determined by the 
energy dispersion, i.e., phonon velocity, of these $\langle1\bar{1}0\rangle$ 
polarized modes. Take the [1$\bar{1}$0] polarized mode for example,  it is
likely to be softer along the [110] direction, i.e., the optic phonon 
propagates with a smaller velocity along the [110] direction; and harder in the
(110) plane. When this mode becomes overdamped at $T_d$, it then 
condenses accordingly, 
with a shorter length scale along the [110] direction than in the (110) plane, 
resulting in the (110) planar correlation of the [1$\bar{1}$0] 
polarizations, and the ``pancake'' shaped entity.

Since PNR are results of soft TO phonon condensation, why do the static diffuse
scattering and  soft optic phonons have different structure factors? This
discrepancy has been very well explained by
Hirota {\it et al.} with the phase-shifted PNR~\cite{PMN_diffuse} model.
They re-examined
the atomic shift values derived from the neutron diffuse scattering measurements
by Vakhrushev {\it et al.}~\cite{PMN_neutron3} on PMN, and found out that 
those values do not satisfy the center of mass condition. The mass center of 
the atoms in the PNR are displaced from the surrounding lattice along the
polarization direction. This displacement is called the ``uniform-phase-shift'',
which is an acoustic type of atomic motion.
The ferroelectric soft phonon mode only condense into the center-of-mass 
portion of the atomic shifts in the PNR, and therefore have different structure
factors than the overall static atomic shifts in the PNR.

In our measurements, we were able to probe diffuse scattering intensities
distributed in three dimensions around many Bragg peaks simultaneously, from 
PZN-$x$PT single crystals. 
The shape of the diffuse scattering in the reciprocal
space can be described as  $\langle110\rangle$ ``rod'' type intensities,
resulting from \{110\} planar correlations of $\langle1\bar{1}0\rangle$ 
polarizations in the real space, with the in-plane correlation length 
about a few to ten times larger than the out-of-plane correlation length.
On the other hand, limited by the measurement resolution, 
we can not accurately determine the values of correlation length in  and out of 
the correlation planes. In addition, we also do not have enough information
to demonstrate whether the in-plane correlations are isotropic, or have
certain preferences. Further studies on these relaxor systems are
required to complete these details and to provide a clear physical picture 
to explain the origin of the planar correlations.

\begin{acknowledgments}
We would like to thank P.~M.~Gehring, S.-H.~Lee, S.~M.~Shapiro, C.~Stock, and 
S.~B.~Vakrushev for stimulating discussions. 
Financial support from the U.S. Department of Energy under contract 
No.~DE-AC02-98CH10886 and the U.S.-Japan Cooperative Neutron Scattering Program 
is also gratefully acknowledged.

\end{acknowledgments}


\end{document}